\documentclass[fleqn,10pt]{wlscirep}
\usepackage[utf8]{inputenc}
\usepackage[T1]{fontenc}

\usepackage{xr-hyper}
\usepackage{hyperref}
\makeatletter
\newcommand*{\addFileDependency}[1]{
  \typeout{(#1)}
  \@addtofilelist{#1}
  \IfFileExists{#1}{}{\typeout{No file #1.}}
}
\makeatother
\newcommand*{\myexternaldocument}[1]{%
    \externaldocument{#1}%
    \addFileDependency{#1.tex}%
    \addFileDependency{#1.aux}%
}
\myexternaldocument{sm}

\title{When Influence Misleads: Informational and Strategic Limits of Social Learning in Trading Networks}

\author[1,4]{Bijin Joseph}
\author[2,4]{Christoph Riedl}
\author[5]{Alex Pentland}
\author[1,4,5,*]{Esteban Moro}
\affil[1]{Department of Physics, Northeastern University, Boston, MA 02115}
\affil[2]{D'Amore–McKim School of Business, Northeastern University, Boston, MA 02115}
\affil[3]{Khoury College of Computer Sciences, Northeastern University, Boston, MA 02115}
\affil[4]{Network Science Institute, Northeastern University, Boston, United States}
\affil[5]{Media Lab, Massachusetts Institute of Technology, Cambridge, MA 02139}

\begin{abstract}
Social learning is a fundamental mechanism shaping decision-making across numerous social networks, including social trading platforms. In those platforms, investors combine traditional investing with copying the behavior of others. However, the underlying factors that drive mirroring decisions and their impact on performance remain poorly understood. Using high-resolution data on trades and social interactions from a large social trading platform, we uncover a fundamental tension between popularity and performance in shaping imitation behavior. Despite having access to performance data, people overwhelmingly choose whom to mirror based on social popularity, a signal poorly correlated with actual performance. This bias, reinforced by cognitive constraints and slow-changing popularity dynamics, results in widespread underperformance. However, traders who frequently revise their mirroring choices (trading explorers) consistently outperform those who maintain more static connections. Building an accurate model of social trading based on our findings, we show that prioritizing performance over popularity in social signals dramatically improves both individual and collective outcomes in trading platforms. These findings expose the hidden inefficiencies of social learning and suggest design principles for building more effective platforms.
\end{abstract}

\usepackage{caption}

\begin{document}

\flushbottom
\maketitle

\vspace{-0.5cm}
\noindent{\small\textsf{\textbf{Keywords:} Social Learning, Popularity Bias, Social Trading, Behavioral Modeling}}

\thispagestyle{empty}

\section*{Introduction}

Collective intelligence and the formation of shared beliefs have been fundamental to the functioning of large, complex societies, with modern networked systems offering unprecedented opportunities for large-scale collaboration \cite{Woolley2010,riedl2021quantifying}. Social learning, defined as the tendency of individuals to emulate the behaviors of others, plays a critical role in facilitating these collective learning processes and is one of the most important behavioral processes in our society and economy \cite{Surowiecki2005,Rendell2010, Pentland2015RealityHedging}. However, the underlying reasons and strategies for social learning in many day-to-day decisions remain elusive. This is partly because it is difficult to observe the motivations behind an individual’s decision to mimic others, and because performance data associated with these behaviors are often absent for many social interactions \cite{riedl2018learning, weng2012competition}. For example, it is difficult to collect data on someone's individual motivation and social influence to visit a specific restaurant and whether they ultimately liked the experience. These limitations hinder a comprehensive analysis of social learning dynamics, which is crucial for understanding the structural properties and evolution of social networks with implications for understanding decision-making, information diffusion, and collective problem-solving  \cite{watts2004new, Lazer2009ComputationalSocialScience}.  

Social learning plays a particularly prominent role in social investment platforms where investors can observe the trades of other investors. Here, investors' decisions are influenced by what they observe others do, and they can sometimes even directly emulate other investors' trading strategies \cite{lo2005adaptive, saavedra2011synchronicity, hirshleifer2023news, bikhchandani1992theory, shiller2000irrational, bouchaud2013crises,Pan2012DecodingSocialInfluence}. Some social trading platforms, such as eToro, ZuluTrade, and Darwinex, allow traders to communicate directly and replicate each other’s trading strategies. As a result, social learning has become increasingly prevalent in these trading platforms. The social trading platform market is projected to expand significantly, growing from approximately USD 2.23 billion in 2021 to USD 3.77 billion by 2028, with a compound annual growth rate (CAGR) of 7.8\% during the forecast period~\cite{insight2025social}. This has driven extensive research focused on the performance and dynamics of social trading platforms.

In social trading platforms, traders can access information about others’ past performance and popularity, and can engage in social learning by allocating funds to mirror selected traders’ strategies. Once set up, the mirroring trader's account will automatically carry out all the trades executed by the chosen trader using the allocated funds, and this will continue until the trader decides to stop. Previous research has explored static potential mechanisms influencing traders' mirroring decisions  \cite{Krafft2016HumanCollectiveIntelligence, almaatouq_adaptive_2020,Dorfleitner2018ToFollowOrNot,krafft2014popularity,liu_coevolution_2023,barkoczi2016social}, as well as platform-wide properties about the dynamics of the social learning \cite{MITpdf2023,somin_remaining_2022}. However, existing literature has overlooked the inherently dynamic nature of social learning, specifically, how traders adjust their mirroring strategies over time in response to evolving performance and popularity signals. Studying this dynamic behavior and its relationship with the rapid-changing nature of financial markets is crucial because traders operate under constraints of limited time, capital, and cognitive capacity, which can severely limit their performance in the market \cite{Miritello2013TimeResource, gigerenzer2001decision, van2010information}. 

Understanding these limitations and the impact of dynamic decision-making strategies on individual traders and trading platforms' collective performance remains an open challenge in social learning research \cite{lo2005adaptive,Rendell2010}. A deeper exploration of these mechanisms could lead to more effective strategies for optimizing trader behavior, enhancing platform performance, and ultimately improving financial outcomes for all participants. 

\begin{figure*}[t] 
    \centering
    \includegraphics[width = \linewidth]{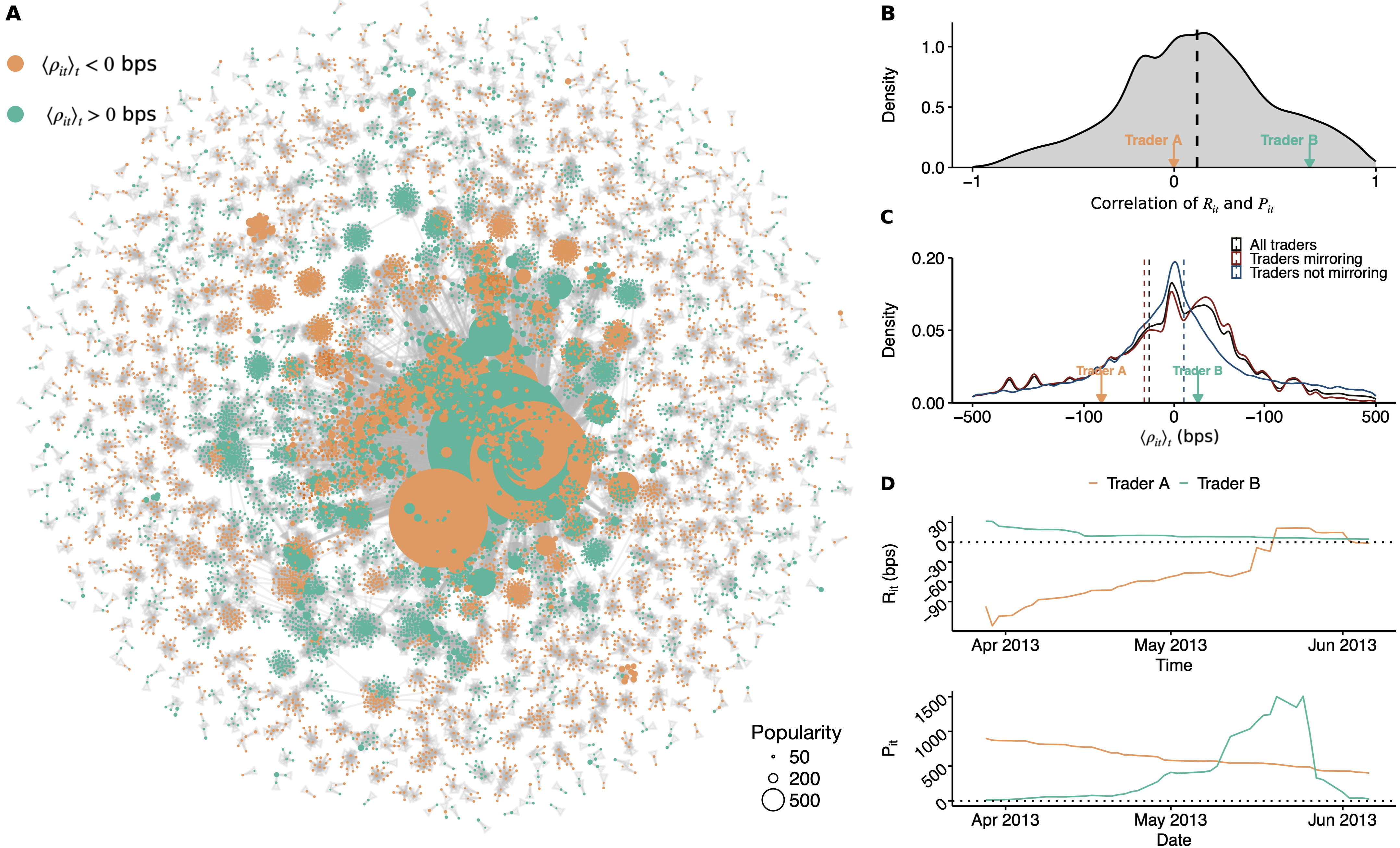}
    \caption{\textbf{Analysis of social learning on social trading platform.} 
A: Network representation of mirroring relationships on eToro. Green nodes represent traders with a positive average performance (\(\langle \rho_{it} \rangle > 0\)) in the observation period $\Omega$, while orange nodes represent traders with a negative average performance (\(\langle \rho_{it} \rangle < 0\)). The size of the nodes indicates their popularity (in-degree).
B: The distribution of the Pearson correlation between the 30-day rolling performance \(R_{it}\) and popularity \(P_{it}\) for different traders (\(i\)) on the platform. The black dotted line represents \(\overline{ \rho_{(R_{it}, P_{it})}} = 0.11 \pm 0.01\).
C: The distribution of the average performance \(\langle \rho_{it} \rangle_t \) of traders on the platform for three groups: for all traders (grey), traders who conduct at least one mirror trade (red), and traders who do not engage in mirror trading (blue). The dotted lines indicate \(\overline{\langle \rho_{it} \rangle_t}\) for each of the three groups.
D: The 30-day rolling performance \(R_{it}\) and popularity \(P_{it}\) of two traders on the platform.  
 }
\label{fig:graph}
\end{figure*}

\section*{Results}

In this study, we investigate the behavioral limitations of social learning and the effects of dynamic decision-making strategies on both individual and collective performance within a trading platform. We do so using a large dataset of social mirroring and trading activity from the eToro social trading platform (see SI Appendix Text~\ref{supp:website}). Over the course of our dataset, the eToro platform's popularity experienced very different phases. Thus, we focus on a time period $\Omega$ during which the trader count was relatively stable, enabling a more consistent study of social learning dynamics (we call this period $\Omega$, details are discussed in SI Appendix Text~\ref{supp:data}). Spanning seven months from Apr 10 to October 16, 2013, the dataset of trades primarily in foreign currency exchange markets made by $164,634$ traders. The dataset includes information on when currencies were bought and sold, the prices at which the currencies were purchased, the number of currency units, the profit made on each trade. It also includes information about whether that trade was done mirroring another trader. About $119,864$ (73\%) of the traders on the platform engage in mirror trading (see SI Appendix Text~\ref{supp:data} for more details about the dataset) while the rest only trade individually. Figure~\ref{fig:graph} shows the network representation of those mirroring relationships in a particular time in $\Omega$. 

During our observation period, the average individual trading performance, defined as the mean Return of Investment (ROI) $\rho_{it}$ across all orders of trader $i$ at different times $t$, is negative, with $\overline{\langle \rho_{it} \rangle_t} = -7.67 \pm 0.41$ basis points (bps; Figure~\ref{fig:graph}C). This means that traders on the platform are, on average, losing money. However, there is a large variability across different groups of traders. In particular, there is a difference between traders who do not engage in mirror trading (i.e., they do only individual trading) and those who do. We find that only $27\%$ traders did not practice mirror trading. This non-mirror group had $\overline{\langle \rho_{it} \rangle_t} = 1.15 \pm 0.94$ bps, while traders using social learning mirroring strategies had $\overline{\langle \rho_{it} \rangle_t} = -10.98 \pm 0.41$ bps. A more detailed individual analysis of mirror trading confirms that this negative effect of social learning persists at the individual level. Specifically, among mirror traders, mirrored trades yield an average performance of $\overline{\langle \rho_{it} \rangle_t} = -61.24 \pm 0.98$ bps, while their individually initiated trades yield a better, though still negative performance of $\overline{\langle \rho_{it} \rangle_t} = -6.88 \pm 0.43$ bps. These findings indicate that underperforming individual traders are more likely to turn to social learning, though this strategy does not substantially improve their outcomes.

\subsection*{Informational limitations} 
Why are traders who engage in social learning on a platform designed for collective performance losing money? Addressing this question requires examining the dynamics of their social learning strategies, their limitations in acquiring and exploiting information on the platform, and their subsequent impact on their performance. First, we analyze the informational signals that traders rely on when initiating or terminating mirroring trades.

On the eToro trading platform, any trader $i$ can access information about any other trader $j$'s popularity, $P_{jt}$, defined as the number of traders mirroring $j$ at time $t$, and their rolling 30-day performance $R_{jt}$, measured as the average return of investment over the last 30 days (see Methods). In principle, in a perfectly efficient and rational market, popularity should reflect performance \cite{lo2005adaptive}.
However, the correlation between $R_{it}$ and $P_{it}$ varies widely (Figure~\ref{fig:graph}C). While some traders exhibit positive or negative correlations between popularity and performance (see Figure~\ref{fig:graph}D), many traders show minimal to no correlation. In SI Appendix Text~\ref{supp:subsec:correlatonPR}, we show that this weak association is not only instantaneous but extends across different time lags, so past performance/popularity does not predict future popularity/performance, respectively.

This lack of correlation can be attributed to multiple factors. Popularity in social networks is significantly influenced by processes like preferential attachment and can be reinforced by visibility biases on the platform, such as ranking algorithms or featured trader lists \cite{barabasi1999emergence, lera2019prediction}. Importantly, popularity evolves slowly over time, as once a trader becomes visible or highly ranked, social reinforcement mechanisms tend to sustain popularity independently of recent performance. On the other hand, we find that traders’ performance is highly volatile and uncorrelated over time, meaning that past performance, which initially drives popularity, is not predictive of future performance (see SI Appendix Text~\ref{supp:subsec:cross-correlation}). As a result, popularity and performance on this platform evolve under largely independent dynamics, leading to a persistent misalignment between social influence and actual financial returns. These dynamics resemble those observed in other social learning platforms \cite{salganik_experimental_2006}, where social influence increases both inequality and unpredictability, and success becomes only weakly tied to intrinsic quality.

We explore the individual origin of this misalignment in more detail by looking at how traders process those signals in their mirroring decisions. To quantify it, we estimate the probability that $i$ mirrors $j$ at time $t$ as  
\begin{equation}
\label{eqLogistic}
P(i\rightarrow j;t) \simeq \mathrm{logit}^{-1}[ \alpha_i + \delta_t + \beta_{\text{per}} R_{jt} + \beta_{\text{pop}} P_{jt}],
\end{equation}
where we account for individual and daily fixed effects. Our results (Table~\ref{tab:model_metrics}) show that traders significantly favor popularity (with log-odds of $\beta_{\text{pop}} = 15.68 \pm 0.58$) over performance ($\beta_{\text{per}} = 0.34 \pm 0.08$) when creating a mirroring relationship, indicating that popularity and thus, preferential attachment, predominantly drives mirroring behavior. A similar pattern emerges when terminating these relationships. Traders disproportionately stop mirroring less popular traders (for more details, see SI Appendix Text~\ref{sup:sec:noderemoval}). These findings suggest that, as mentioned before, traders prioritize social signals of popularity rather than objective performance metrics, leading to the misalignment of social signals and, thus, diminishing their informational value. This misalignment may explain why traders relying on social learning tend to lose money on average, as their decisions are driven by social influence derived from signals about popularity rather than signals of financial performance.

\begin{table}[t]
\begin{center}
\begin{tabular}{l c}
\hline\hline
$\beta_{\text{per}}$ (performance)    & $0.34^{***}$   $(0.02)$      \\
$\beta_{\text{pop}}$ (popularity)    & $15.68^{***}$   $(0.58)$      \\
\hline
\textit{Fixed-effects} \\
day & yes \\
traders & yes \\
\hline
\textit{Fit statistics} \\
Observations                   & $160890$     \\
Squared Correlation            & $0.57$        \\
Pseudo $R^2$                    & $0.29$        \\
Accuracy                       & $0.83$        \\
Accuracy $95 \% $ CI           & [$0.82, 0.83$] \\
\hline\hline
\multicolumn{2}{l}{\scriptsize{$^{***}p<0.001$; $^{**}p<0.01$; $^{*}p<0.05$}}
\end{tabular}

\caption{\textbf{Results of the logistic regression model}.The estimated coefficients of the logistic regression model (\ref{eqLogistic}) and their respective standard errors in parentheses. The p-values
correspond to two-sided tests for the hypothesis that each coefficient is different from zero. We also
report the Squared Correlation and Pseudo $R^2$
results for the logistic regression, and the accuracy in the training of the model and its 95 \% confidence interval in square brackets.}
\label{tab:model_metrics}
\end{center}
\end{table}

\subsection*{Dynamic strategy limitations}
\begin{figure*}[t!] 
    \includegraphics[width = \linewidth]{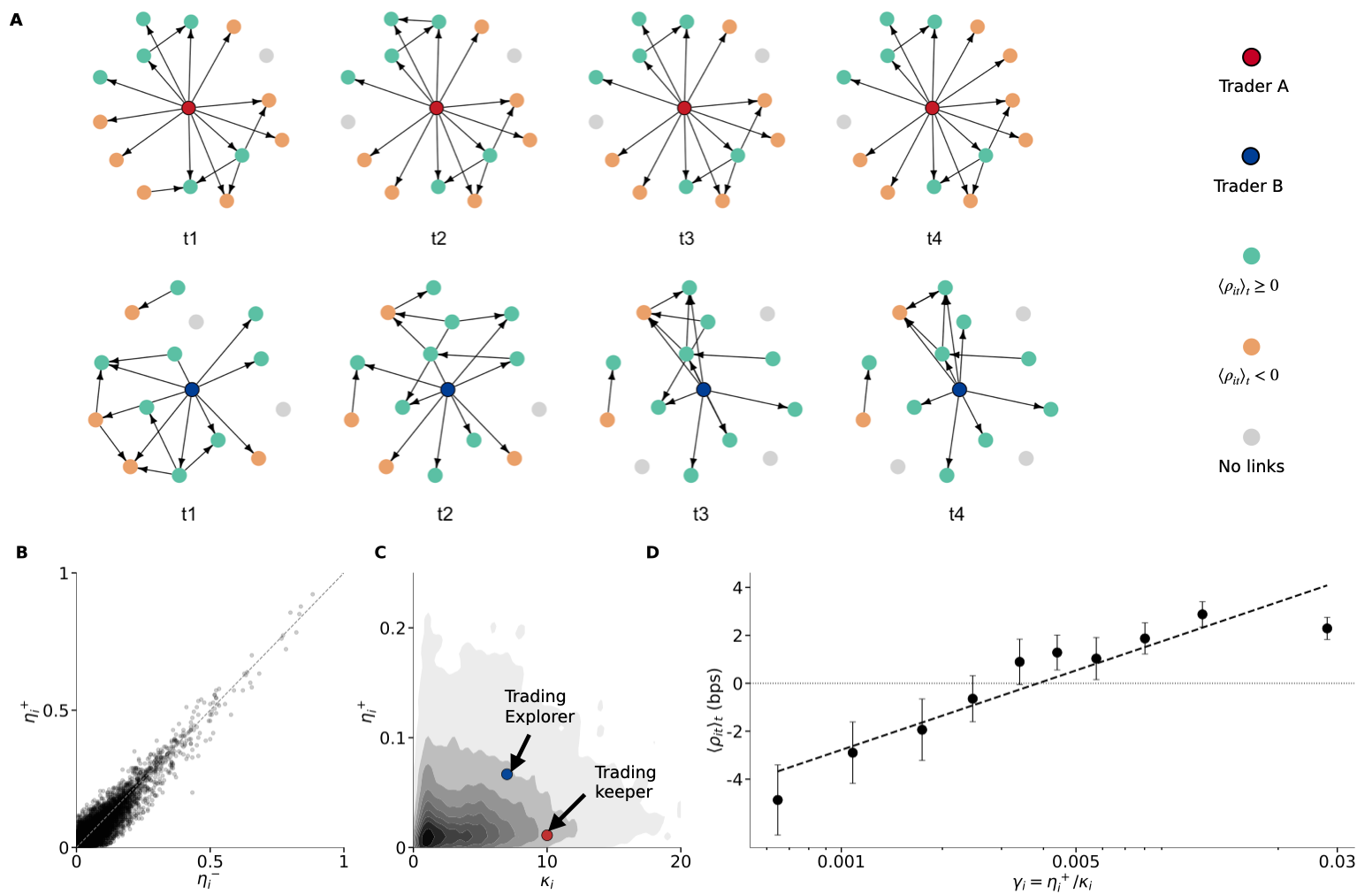}
    \caption{\textbf{Dynamic strategies on the social trading platform.} 
A: Temporal network representation of the mirroring relationships between two traders (Trader A and B) on the platform across four equally distant times in $\Omega$.  
B: Comparison between the total number of mirrors added, $\eta_i^+$, and the total number of mirrors removed, $\eta_i^-$ for each trader $i$ during $\Omega$.
C: Comparison between $\eta_i^+$ and $\kappa_i$, where $\kappa_i$ represents the capacity, or the number of active mirrors maintained by trader $i$. The red marker represents trader A, and the blue marker represents trader B in panel A.  
D: Binned scatter plot between average performance, $\langle \rho_{it}\rangle_t$, and $\gamma_i = \eta_i^+ / \kappa_i$. Here, $\gamma_i$ represents the rate at which a trader changes their social mirrors. The dashed line shows a linear-log fit with slope = $2.06 \pm 0.28$. (for more information see see SI Appendix Table~\ref{supp:table:S5_regression_panelC}).}\label{fig:comparison}
\end{figure*}

While traders primarily rely on popularity for social learning, Equation (\ref{eqLogistic}) shows that performance metrics still play a small role in their decision-making. Thus, a rapid adaptation strategy in which traders frequently adjust mirroring relationships could eventually allow them to align with profitable strategies even if performance is only weakly used. However, our findings indicate that most traders are not employing such an adaptive approach, likely due to cognitive, temporal, or financial limitations \cite{lo2005adaptive}.
If $\kappa_{it}$ is the number of mirrors maintained by trader $i$ at time $t$, we find that, on average, these traders maintain $\overline{\langle{\kappa_{it}}\rangle_t} = 2.26 \pm 0.01$ simultaneous mirrors. Furthermore, traders continuously modify their mirroring trades as we can see in Figure~\ref{fig:comparison}A. To quantify the dynamics of this strategy, we introduce two metrics: the mirror rate gain ($\eta_i^+$), representing the daily rate of creation of new mirrors an agent $i$, and the daily mirror rate loss ($\eta_i^-$). Our findings reveal a strong correlation between those rates (Figure~\ref{fig:comparison}B), indicating that traders balance adding and removing mirrors. This implies that, despite external influences and available information and changing conditions in the market, each trader maintains a relatively stable number of simultaneous active mirrors at any given time, $\kappa_{it} \simeq \kappa_i$.

The existence of a finite social learning capacity is consistent with findings from other social networks and human activities, which show that individuals are limited in the number of relationships or tasks they can manage simultaneously \cite{Miritello2013TimeResource, dunbar1998social, andrea}. In the trading platform, each trader manages those restrictions with a well-defined dynamic strategy to maintain and revise mirroring trades. That social trading strategy is defined by $\kappa_i$, the number of mirrors maintained instantaneously, and $\eta_i^+$, the ratio at which mirrors are created/removed. As shown in Figure~\ref{fig:comparison}C, there seems to be no relationship between both metrics, suggesting significant variation in traders' dynamic strategies for adaptive social learning. On one end, traders with large $\gamma_i = \eta_i/\kappa_i$, referred to as {\em trading explorers} \cite{Miritello2013LimitedCommunication}, frequently revise and change their mirroring relationships. Conversely, {\em trading keepers} exhibit low $\gamma_i$, maintaining a stable set of mirroring throughout our period $\Omega$. 

This variability in social learning strategies has a clear impact on the performance of traders. We find a positive correlation between $\gamma_i$ and average trader performance $\langle \rho_{it} \rangle_t$ (Figure~\ref{fig:comparison}D). This result indicates that even in a rapidly evolving trading platform, where past performance is not a reliable predictor of future success and people are using mostly popularity vs.~performance, traders who frequently adjust their mirroring relationships (explorers) are still able to profit from their social learning strategies. On the contrary, most traders rarely update their mirroring relationships, leading to suboptimal adaptation to market changes and individual and platform-wide losses.

\par

\subsection*{Modeling efficient social trading platforms}
Our findings highlight the crucial interplay between informational (popularity vs.~performance) and cognitive limitations in social learning. Understanding and managing this balance is essential for improving individual performance and optimizing the individual and overall efficiency of social trading. To explore how more efficient social trading platforms could be designed, we develop a model of social learning informed by insights from our analysis of the eToro dataset. Our model represents the trading platform as a directed temporal network where nodes correspond to traders and edges represent mirroring relationships, as in Figure~\ref{fig:graph}A. The edges' direction indicates the mirroring relationship's direction from mirroring trader $i$ to mirrored trader $j$. Mirrors are created according to equation (\ref{eqLogistic}) using fixed values of $\beta_{\text{per}}$ and $\beta_{\text{pop}}$ to control the relative influence of performance and popularity in imitation behavior across the platform. Varying these values allows us to explore how different emphasis on social versus performance signals affects collective dynamics, and individuals' and the platform's overall efficiency. Consistent with our findings, each trader in our simulation is defined by a dynamic strategy $(\kappa_i,\eta_i^+)$ for their mirroring. At each trading day $t$, a trader $i$ starts a mirroring relationship with a rate proportional to $\eta_i^+$ and chooses a node $j$ to mirror with probability given by equation (\ref{eqLogistic}). 

After each trading day $t$, the return of investment of agent $i$, is calculated as follows:
\begin{equation}
\rho_{i,t} = \frac{1}{\kappa_{i,t-1}} \sum_{j} A_{ij,t-1} \rho_{j,t-1}  + \epsilon_{it}.
\end{equation}
Here, the first term on the right-hand side represents performance from their social trading, i.e., the aggregated performance of all traders mirrored by $i$ at a previous day $t-1$, while $\epsilon_{it}$ accounts for independent trading returns for trader $i$ at day $t$. The adjacency matrix $A_{ij,t-1}$ captures the evolving structure of mirroring relationships between traders and, again, $\kappa_{i,t-1}= \sum_{j} A_{ij,t-1}$ denotes the number of mirrors maintained by trader $i$ at time  $t-1$. Note that in our simulation, we assume that the amount of money placed by $i$ on each mirror is constant and equally distributed to $1/\kappa_{i,t-1}$. We also assume that traders invest equally between their social learning strategy and independent trading. To simulate the independent trading performance $\epsilon_{it}$, we use an Ornstein-Uhlenbeck (OU) process \cite{Maller2009}, which is the simplest stochastic process with temporal correlation and reflects the finding that traders can have independent strategies with short time-correlated periods of positive or negative performance. The OU process parameters were calibrated to reproduce individual performance in the eToro platform (see Methods and SI Appendix Text~\ref{supp:sec:model} for more details about our model). 

\begin{figure*}[t] 
    \centering 
    \includegraphics[width = \linewidth]{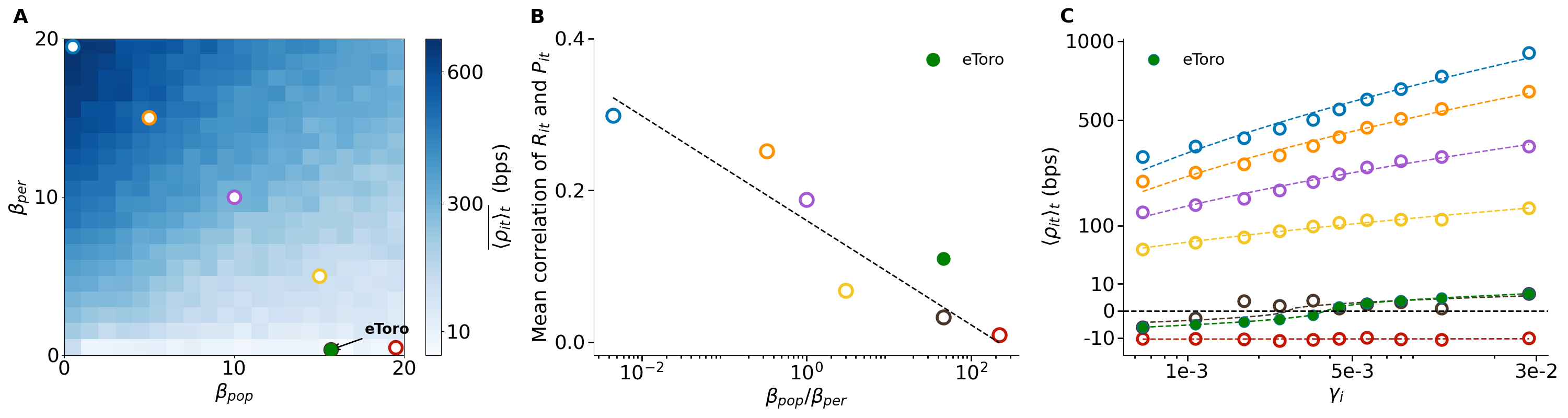}
    \caption{ \textbf{Modeling efficient social trading platforms.} 
A: Platform-wide performance in our simulations as a function of how much focus users in the platform put on popularity $\beta_{\text{pop}}$ or performance $\beta_{\text{per}}$ in making a mirroring decision. 
B: Change in the alignment of social signals in our simulations (measured as the Pearson correlation between $R_{it}$ and $P_{it}$) for different platform conditions, $\beta_{\text{per}}$ and $\beta_{\text{pop}}$ values in our simulations. Dashed line is a linear-log fit to the data.
C: Relationship between individual average performance and the dynamical strategy (represented by $\gamma_i$) for different platform conditions. Dashed lines are fits to each condition, see SI Appendix Table~\ref{supp:table:S5_regression_panelC}.}\label{fig:heatmap}
\end{figure*}

To test the validity of our model, we compare its outcomes with empirical results from the eToro platform using the values $\beta_{\text{pop}}$ and $\beta_{\text{per}}$ in Table~\ref{tab:model_metrics}. Despite its simplicity, our model describe quite accurately the main empirical findings in eToro. For example, the average performance in the simulation, $\overline{\langle \rho_{it} \rangle}_t = -16.97 \pm 1.81$ bps, is close to that of social learners (those who engage in mirroring trading) on eToro, which is $-10.98 \pm 0.41$ bps. The Pearson correlation between $R_{it}$ and $P_{it}$ in the model is $0.05 \pm 0.01$, compared to $0.11 \pm 0.01$ observed on eToro. 
As shown in SI Appendix Text~\ref{sup:subsec:corrR} and~\ref{supp:subsec:cross-correlation}, the dynamic properties of popularity and performance show similar patterns in both the empirical data and the model. Additionally, we found that a more-adaptive rapid mirror strategy leads to better individual outcomes. In particular a linear regression of $\langle \rho_{it} \rangle_t$ as a function of $\gamma_i$ yields slopes of $1.54 \pm 0.12$ in the model, compared to $2.06 \pm 0.28$ in the eToro data (See Figure~\ref{fig:comparison}D and SI Appendix Table~\ref{supp:table:S5_regression_panelC}). These similarities suggest that our model incorporates the right ingredients to reproduce social learning and the impact of informational and cognitive limitations on individual and platform performance.

Modifying the relative emphasis on popularity $\beta_{\text{pop}}$ and performance $\beta_{\text{per}}$ in our model allows us to simulate the effect of those signals' influence and network dynamics on the individual and global performance in other potential social trading platforms. Figure~\ref{fig:heatmap}A shows how the average ROI of all traders varies with the relative strength of the signals used to mirror in the platform. As we can see, emphasizing performance over popularity leads to a marked increase in the platform's performance in general. However, due to eToro’s trader behavior and design choices, the platform ends up in one of the least effective regions of the performance–popularity space (Figure~\ref{fig:heatmap}A). For example, reducing the ratio of popularity to performance used by users in their mirroring decisions by $10\%$ would increase overall performance by $6.6\%$. As shown in Figure~\ref{fig:heatmap}B, that improvement comes from the increased alignment between popularity and performance signals when traders on the platform emphasize performance in their mirroring decisions. 

In all cases, we also find that dynamic social learning strategies consistently outperform static ones (see Figure~\ref{fig:heatmap}C). Traders with larger $\gamma_i$, who regularly seek new individuals to follow, achieve better outcomes than those who maintain stable mirroring relationships. This advantage becomes especially pronounced when performance signals are prioritized ($\beta_{\text{pop}}/\beta_{\text{per}} \ll 1$), as active explorers continuously identify and follow traders with higher performance. As a result, traders with larger $\gamma_i$ can get up to 3x larger performance than those with smaller $\gamma_i$.

\section*{Discussion}

Behavioral and structural limits of social learning can hugely impact the ability of humans to benefit from it \cite{banerjee1992simple, gigerenzer1996reasoning, giraldeau2002potential, mason2012collaborative}. Our study reveals how these limitations manifest in a large social trading platform, where traders can imitate others based on visible signals of both popularity and performance. Despite this, most users tend to follow popular traders, not the most successful ones—leading to widespread underperformance. This result aligns with other findings in product \cite{salganik_experimental_2006} and  financial markets \cite{barber2008all}, who show that individual investors are more likely to be influenced by attention-grabbing products or stocks than by fundamentals, often to their detriment.

We identify that the key mechanism is the constant misalignment between social signals on the platform. This misalignment likely stems from a combination of popularity reinforcement dynamics (e.g., preferential attachment), platform-level choices about how information is displayed, and user cognitive limitations to process rapidly changing data. Yet, even in the presence of this structural mismatch, some users consistently outperform others. Traders who adopt more adaptive strategies—frequently updating whom they mirror—can better navigate the noisy information environment and mirror high-performing peers. This suggests that rapid adaptability in social learning strategies can partially compensate for signal misalignment and lead to better individual outcomes. Platforms or systems that enable users to regularly revisit and revise their social learning strategies may allow them to overcome structural processes based on popularity, which otherwise would dominate over longer timescales \cite{salganik_experimental_2006}. Finally, rapidly evolving strategies might also help mitigate the cost of herding in those social trading platforms by breaking trading bubbles around the most popular investors \cite{almaatouq_adaptive_2020, Pan2012DecodingSocialInfluence}.

Our study has some limitations. For example, traders pursue different goals with some favor strategies that minimize risk and provide steadier returns, while others are willing to accept higher risks in pursuit of greater rewards \cite{barberis2003survey}. In this study, we focused only on maximizing returns, regardless of risk. Incorporating heterogeneous utility functions or bounded rationality might offer a more realistic picture of social learning in decision-making under uncertainty \cite{lo2005adaptive}. Additionally, our analysis covers a relatively short stable time frame. Social learning dynamics might evolve differently over longer periods, or under extreme conditions like market crashes \cite{saavedra2011synchronicity}. Yet, the fact that informational and dynamical limitations are common across human social learning systems—and that our simple model can successfully account for the patterns observed in the platform—suggests that our results may be generalizable to address those limitations.

In an increasingly interconnected and fast-changing world, our findings hold broader significance when it comes to engaging with friends, celebrities, influencers, or politicians \cite{muchnik2013social, loader2016performing, van_zoonen_entertaining_2005, shamim2024power, barquero2023owned}. The misalignment between social signals and actual performance that we observe in trading platforms extends well beyond the domain of finance. Similar phenomena characterize other digital environments where popularity metrics—such as likes, views, or follower counts—heavily influence user behavior, often at the expense of content quality or informational accuracy. This phenomenon has been documented in the spread of misinformation on social media \cite{bucher_want_2012}, the amplification of polarizing content in political discourse \cite{vosoughi_spread_2018}, and the unfair over-representation of particular products in entertainment platforms \cite{abdollahpouri_connection_2020,salganik_experimental_2006}. Future work should explore how these social learning dynamics interact with varying market structures, algorithmic recommendation systems, and behavioral heterogeneity to better inform both theory and design in socially-mediated decision environments. Designing systems that support adaptive social learning, while mitigating the distortive effects of popularity-based signals, will be essential for fostering fairer and more effective collective outcomes.

\section*{Methods}

\subsection*{Dataset} \label{Dataset}
Our dataset was obtained from the eToro company (etoro.com), an online retail broker for foreign exchange, index, and commodities trading, and one of the largest social trading platforms. The dataset contains $87.5$ million anonymized trades and $825$k mirroring relationships from August 2010 to December 2013. The dataset was preprocessed and prepared in \cite{krafft2014popularity}. As the platform gained popularity during its initial growth phase, the number of users increased significantly. To study trader social learning behavior during a more stable period, we focus our analysis on the time interval denoted as $\Omega$—from April 2013 to October 2013—when the daily number of traders doing mirroring trading stabilized around 50k (for more details, see SI Appendix Text~\ref{supp:data}).

\subsection*{Definition of performance and popularity} 
In our analysis, we only considered realized profits. When a trader $i$ closes her trade at time $t$, the performance of that trade is $\rho_{it}$. The mean performance of $i$ across $\Omega$, $\langle \rho_{it} \rangle_t$ is the mean of $\rho_{it}$ for all trades where $ t \in \Omega$. Figure~\ref{fig:graph}C displays the population diversity of $\langle \rho_{it} \rangle_t$ for all traders, while values on the text are the average values of them over all traders, $\overline{\langle \rho_{it}\rangle_t}$. On the platform, traders are provided with a rolling 30-day performance $R_{it} = \frac{1}{N_{it}}\sum_{\tau \in [t-30,t]} \rho_{i\tau}$ for every trader, where $N_{it}$ is the number of trades closed by $i$ in the last 30 days. Finally, $P_{it}$, the popularity of user $i$ at time $t$ is reconstructed by getting all mirroring relationships to $i$ at that time. For more details on who we reconstruct and calculate trade, individual performance and popularity, see SI Appendix Text~\ref{supp:data}.

\subsection*{Mirror prediction}
Before making a mirroring decision, a trader \( i \) has access to \( R_{jt} \) and \( P_{jt} \) of other traders \( j \) on the platform. To estimate the probability $P(i \to j;t)$ in Equation~(\ref{eqLogistic}), we take all events in which a trader $i$ creates a mirror relationship with $j$ at time $t$ as positive examples. For each event $i\to j$ at time $t$ we select another trader $j'$ (not mirrored by $i$ at time $t$) and take $i \to j'$ as a negative example. Thus our training sample of mirroring events is balanced. We normalize both $R_{jt}$ and $P_{jt}$ across the training set before estimating the parameters $\beta_{\text{per}}$ and $\beta_{\text{pop}}$.

\subsection*{Modeling Social Learning} 
\label{modelling}
\subsubsection*{Mirror dynamics}
In our model, each agent is characterized by a dynamical mirroring strategy defined by ($\kappa_i$,$\eta_i^+$). At any time step, trader $i$ decides to mirror another trader $j$ with rate $\eta_i^+$ and chooses $j$ using with a probability $P(i\to j;t) \sim \text{logit}^{-1} [\beta_{\text{per}} R_{jt} + \beta_{\text{pop}} P_{jt}]$ 
similarly to Equation\ (\ref{eqLogistic}) found for the real data. Note that, for simplicity, in our model, we do not consider individual preferences or daily effects in the agent's mirroring decision. To maintain a stable number of simultaneous active mirrors, after $i$ creates a new mirror, we remove another $i \to k$ previously formed mirrors with the lowest $P(i \to j)$. See more details about our dynamical social learning model in SI Appendix Text~\ref{supp:sec:model}.

\subsubsection*{Simulating independent investments} \label{methods:OU}
In our study, we simulate independent investments made by the traders $\epsilon_{it} $ using an Ornstein-Uhlenbeck (OU) process. The OU process captures the idea that traders, while employing independent strategies, exhibit short-term periods of correlated positive or negative performance that tend to revert to a long-term mean. To reflect the dynamics of the platform, we calibrated the OU process to reproduce the aggregated time structure and short auto-correlation of individual performance (for more details, see SI Appendix Text~\ref{supp:data:OU}).

\section*{Data and Code Availability}
Data and code to recreate the Figures and Table are available on GitHub.\\
\href{https://github.com/SUNLab-NetSI/when-influence-misleads}{https://github.com/SUNLab-NetSI/when-influence-misleads}.


\section*{Acknowledgements}

We thank Yaniv Altshuler and P.M. Krafft for providing the dataset and for discussions about the data. We thank Hamish Gibbs, Lucy Butler, Guangyuan Weng, Saumitra Kulkarni, and Marco Tonin for their valuable insights on the paper. E.M. and B.J. acknowledge support from the U.S. National Science Foundation under Grants 2420945 and 2427150. 

\section*{Author contributions statement}

B.J. and E.M. designed and performed research; All authors supervised research, analyzed the results, and co-wrote the manuscript.

\section*{Ethics declarations}

\subsection*{Competing interests} 
The authors declare no competing interest.

\clearpage
\section*{Supplementary Information}

\section{Social Trading Platforms} \label{supp:website}
A social trading platform is built on the idea of social learning--the notion that the collective wisdom of many traders can outperform the insights of a single individual. Platforms like \textit{ZuluTrade}, \textit{eToro}, and \textit{Darwinex} leverage this concept by allowing users to trade independently or follow and copy the strategies of other successful traders. But this idea extends far beyond trading---it also influences how people interact on social media, citation networks, streaming platforms, and more. That’s why understanding the dynamics within these systems is both important and necessary.
\begin{figure}[h]
    \centering
    \includegraphics[width=\linewidth]{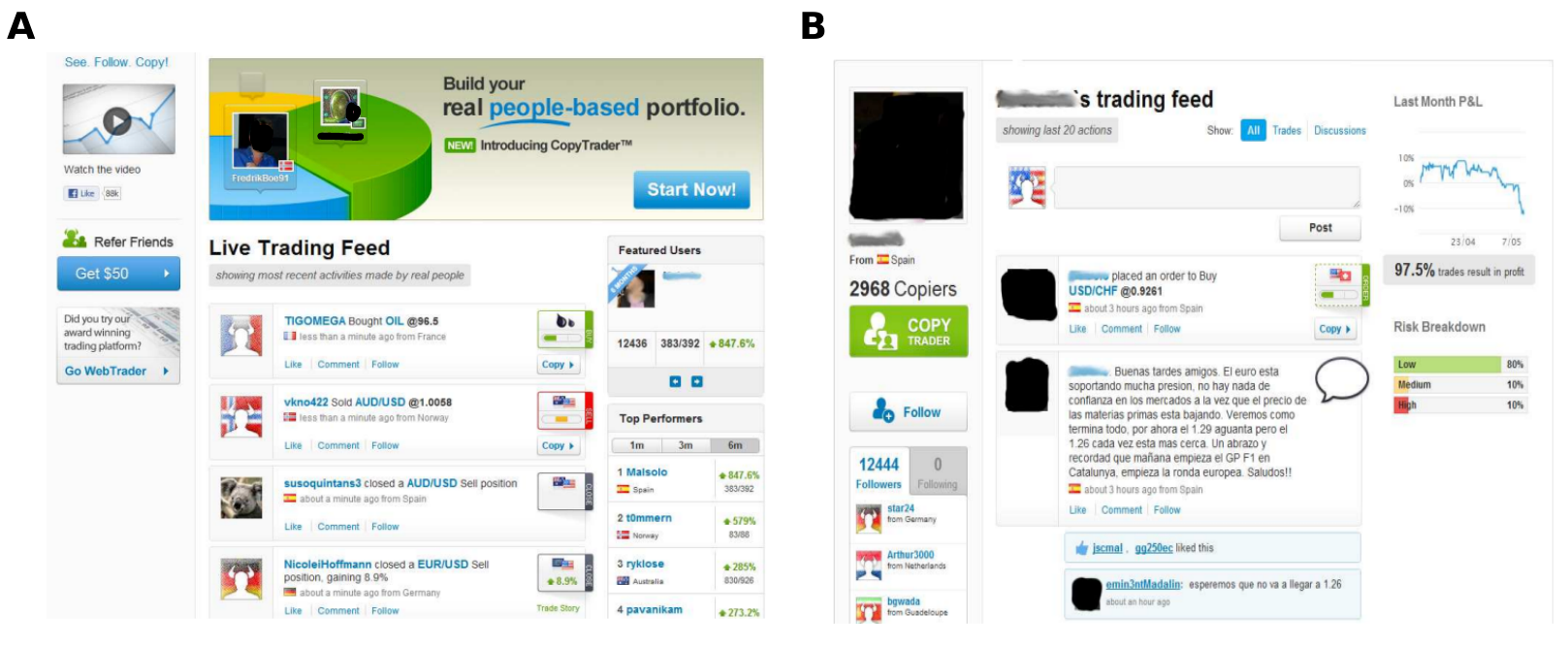}
    \caption{\textbf{The eToro social trading platform.} A: The general landing page shows the current trades by other users and top-ranked traders. Users can click any trade to copy. B: Public profile page for a eToro user (images and names removed), which contains his current trades, messages, and most importantly, the number of followers ({\em copiers}) mirroring his trades. Image taken from \cite{Pan2012DecodingSocialInfluence}. }
    \label{supp:fig:platform}
\end{figure}

In this study, we used data from eToro (\texttt{etoro.com}) to examine the dynamics of social learning. eToro is one of the largest social trading platforms, operating as an online retail broker for foreign exchange, indices, and commodities trading. At the time of our data collection, the platform primarily offered two core functionalities: (1) enabling users to execute their own trades, and (2) allowing users to discover and replicate the trades of other traders on the site, see Figure~\ref{supp:fig:platform}. Notably, eToro imposed no cap on leverage, meaning users could potentially incur losses exceeding 100\% of their position value in a single transaction. In addition, the platform featured live streaming feeds that included fundamental announcements, market news, trading activity from fellow users, and insights from top-performing traders. While eToro’s internal algorithms and website interface have evolved over time, the features described above accurately characterize the platform during the period covered by our dataset.

During the period of our data collection, eToro was primarily active in the foreign exchange market, with most trading activity centered around currency pairs—that is, the buying of one currency while selling another. In addition to providing a platform for individual trading, it featured a social learning tool known as OpenBook, which allowed users to view the historical performance and popularity of other traders. Users could participate in three distinct modes of trading: executing single trades based on their own decisions, copy trading by replicating specific trades made by others, and mirror trading, where they could select a trader to follow and have eToro automatically execute all future trades made by that trader on their behalf. \par

\par
\begin{figure}[t]
    \centering
    
    \includegraphics[width=10cm]{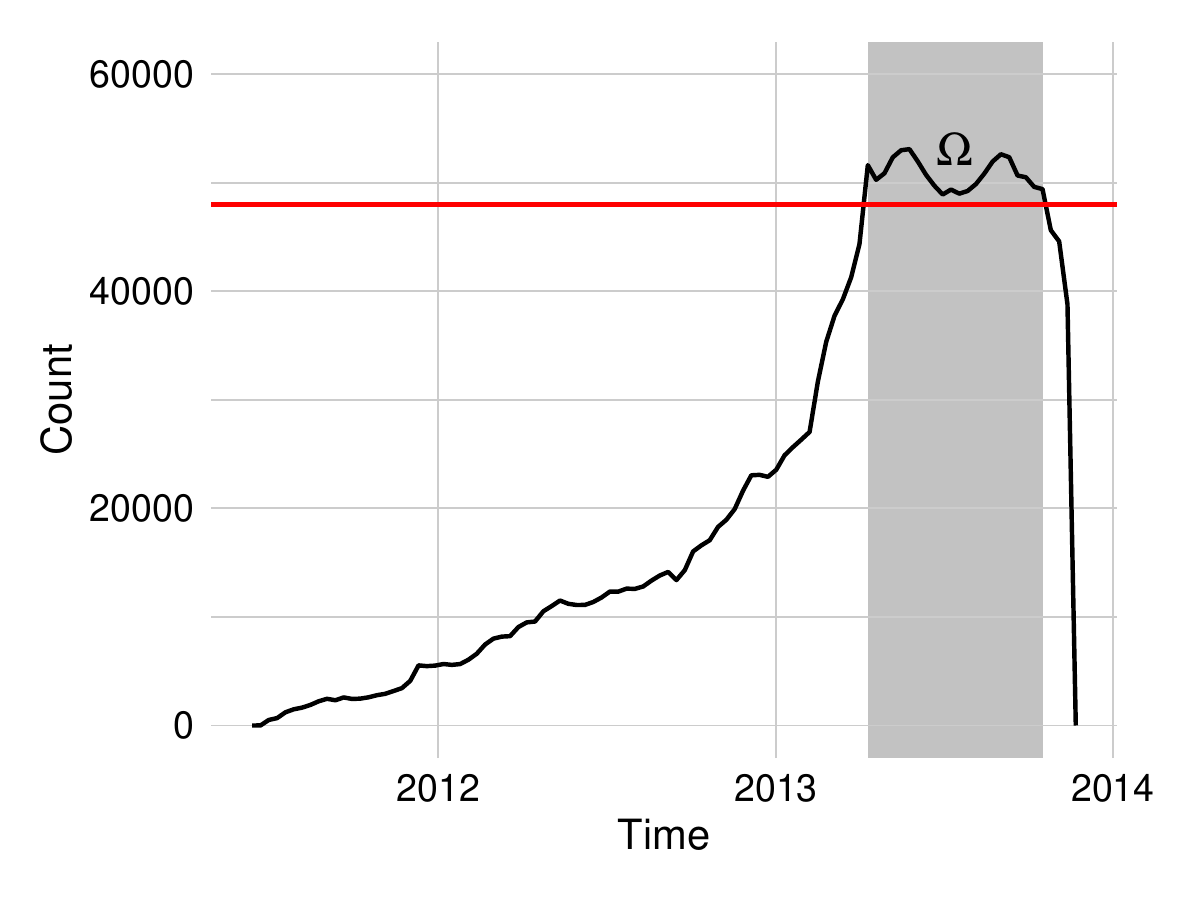}
    \caption{\textbf{Time series of the number of users mirroring on the eToro platform.} The daily number of traders mirroring on the platform increases from 2011-08 to 2013-03 after which it fluctuates around $50{,}000$. After October 2013, the number of users decreases.  The stable shaded period, denoted as $\Omega$, is from April 2013 to October 2013. }
    \label{supp:fig:omega}
\end{figure}

\section{Data} \label{supp:data}

Our dataset comes from the company eToro and consists of a set of trades from the eToro website. The dataset was preprocessed and prepared in \cite{krafft_bayesian_2021,krafft2016rational}. The dataset contains two tables, one for each trade (top) and one for each of the mirroring relationships (see Table~\ref{supp:data:table:trades}). Each trade is identified by the id of the trader placing that order {\tt CID}, the id for the trader that originated that order {\tt parentCID} (if that trade is mirrored, otherwise 0), the time the order was opened/closed {\tt openDate/closeDate}, the volume of currency traded {\tt unitsDecimal}, the open rate for the traded currency {\tt openRate}, and the net profit from the trade {\tt nProfit}, see \cite{krafft_bayesian_2021}.

\par
\begin{table}[t]
\centering
\begin{tabular}{rrlrlrrl}
\toprule
{\tt CID} & {\tt openDate} & {\tt parentCID} & {\tt unitsDecimal} & {\tt openRate} & {\tt nProfit} & {\tt closeDate} \\
\midrule
2139256 & 2013-04-09T16:22:37Z & 0 & 1000.000000 & 1.308500 & 0.600000 & 2013-04-10 \\
2635717 & 2013-04-09T16:33:10Z & 2635681 & 1000.000000 & 0.851600 & 0.500000 & 2013-04-10 \\
2003261 & 2013-04-09T08:31:50Z & 0 & 27.000000 & 1.529200 & 0.080000 & 2013-04-10 \\
2687374 & 2013-04-09T15:51:33Z & 2130217 & 10000.000000 & 1.308700 & 9.000000 & 2013-04-10 \\
2100958 & 2013-04-09T23:59:23Z & 0 & 10000.000000 & 98.950000 & -4.040000 & 2013-04-10 \\
... & ... & ... & ... & ... & ... & ... \\ 
\bottomrule
 &  &  &  &  &  &  \\ 

\end{tabular}

\begin{tabular}{rrrll}
\toprule
{\tt CID} & {\tt parentCID} & {\tt mirrorID} & {\tt openDate} & {\tt closeDate} \\
\midrule
692486 & 273268 & 1 & 2011-06-13T08:33:05Z & 2011-06-13T08:34:02Z \\
173764 & 139841 & 11 & 2011-06-14T13:26:14Z & 2011-06-14T14:42:43Z \\
436771 & 388516 & 17 & 2011-06-15T08:56:13Z & 2011-06-15T10:54:57Z \\
180610 & 1553412 & 38 & 2011-06-19T21:03:00Z & 2011-06-19T21:05:45Z \\
194221 & 180610 & 53 & 2011-06-21T09:11:51Z & 2011-06-21T14:30:39Z \\
... & ... & ... & ... & ... \\
\bottomrule
\end{tabular}
\caption{\textbf{Top: Trades dataset.} In the trades dataset, {\tt openDate} and {\tt closeDate} indicate when a currency trade was opened and closed, respectively, by trader {\tt CID}. The {\tt parentCID} id identifies the trader that originated that order if it was mirrored (otherwise it is 0). The trade volume is given by {\tt unitDecimals}, {\tt openRate} is the initial currency price, and {\tt nProfit} is the resulting profit, in the same currency type. \textbf{Bottom: Mirror dataset.} In the mirror dataset, {\tt CID} identifies the trader executing the order, and {\tt parentCID} refers to the mirrored trader's {\tt CID}. Each mirroring trade has a unique {\tt mirrorID}, with {\tt openDate} and {\tt closeDate} indicating the start and end of the mirroring order, respectively.}
\label{supp:data:table:trades}
\end{table}

The mirror dataset (Table~\ref{supp:data:table:trades}) only contains the information about every mirroring relationship which is described by the id of the trader mirroring {\tt CID}, the trader mirrored {\tt parentCID}, and opening/closing dates for that relationship {\tt openDate/closeDate}.

The dataset spans the period between June 2011 and November 2013, with a total of 87.9 million recorded trades by 340k traders and 188k mirroring relationships. However, as shown in Figure~\ref{supp:fig:omega}, the platform went through a number of changes during the first years, especially in the fraction and number of users that engaged in social learning. As shown in Figure~\ref{supp:fig:omega}, the number of traders that mirrored increases from August 2011 to around March 2013, after which it remains relatively constant until October 2013. Following this period, the number of users drops sharply because data collection was stopped.  Contrary to other studies that focused in the initial exponential growing phase \cite{krafft_bayesian_2021}, in our analysis, we focus only on the time window where the platform reached maturity and stability, during which the number of traders engaging in social learning on the platform remains constant (approximately 50,000). This selected period from March 2013 to October 2013 is highlighted in Figure~\ref{supp:fig:omega} by the shaded region, denoted as $\Omega$. During the period $\Omega$, we have about 190,351	traders in the dataset. 

\begin{figure}[t]
\centering

\begin{minipage}{.5\textwidth}
  \centering
  \makebox[-220pt][r]{\raisebox{1.0 em}{\hspace{6.0cm}\textsf{\textbf{\LARGE A}}}}  

  \footnotesize
  \label{table-label0}
    \caption*{Power-law fit statistics across dates}
\begin{tabular}{lcccc}
\toprule
Date & 2013-05-20 & 2013-07-09 & 2013-08-28 & 2013-10-17 \\
\midrule
$\beta$ & 1.61**  & 1.61**  & 1.60*  & 1.60*  \\
        & (0.02) & (0.02) & (0.02) & (0.02) \\
$x_{min}$ & 2.00 & 2.00 & 3.00 & 3.00 \\
\hline
$\mathrm{Observations}$ & 1736 & 1521 & 1164 & 1070 \\
$\log L_{pl}$ & -6124.56 & -5386.37 & -4750.80 & -4382.19  \\
$\log L_{ln}$ & -6121.42 & -5383.02 & -4747.74 & -4379.89 \\
$R$ & -3.14 & -3.35 & -3.06 & -2.29 \\

\bottomrule
\end{tabular}

\vspace{0.5em}
\noindent\footnotesize\textit{Note:}  $^{*}$p$<$0.1; $^{**}$p$<$0.05; $^{***}$p$<$0.01 \\
\vspace{5.5em}
\end{minipage}%
%
\begin{minipage}{.5\textwidth}
  \centering
  \makebox[-210pt][r]{\raisebox{-1.5em}{\hspace{6.6cm}\textsf{\textbf{\LARGE B}}}}  

  \vspace{1.5em}
  \includegraphics[width=\linewidth]{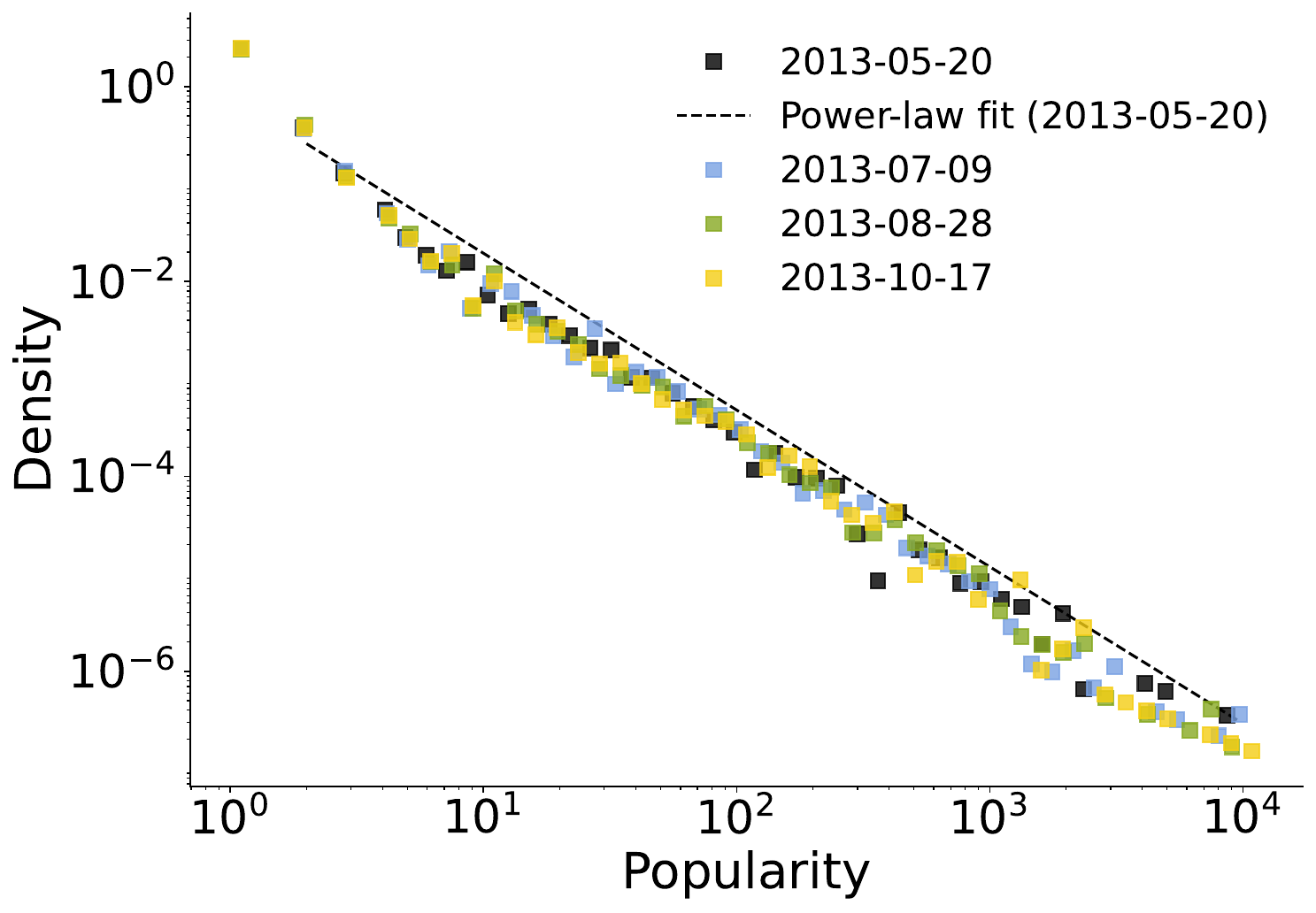}
\end{minipage}
\caption{\textbf{Distribution of popularity on the platform.} A: Results of fitting the popularity distribution to a power-law model $P(P_{it}) = A P_{it}^{-\beta}$ using the package {\tt powerlaw} \cite{power-law} The table reports the estimated coefficients along with their corresponding standard errors (in parentheses). P-values from two-sided tests are included to assess whether each coefficient significantly differs from zero. The $\log L_{pl}$ is the log-likelihood of the data under a power-law distribution, $\log L_{ln}$ is the log-likelihood of the data under a log-normal distribution and $R = \log L_{pl} - \log L_{ln}$ is the likelihood ratio. B: The graphical representation of the popularity distribution for different time periods.}
\label{supp:fig:data-label}
\end{figure}

During $\Omega$, the system exhibits a dynamically stable temporal network $A_{ij,t}$. This can be seen in the macroscopic network metrics. In that network, the out-degree of a trader $i$ at time $t$ represents the number of active mirror connections initiated by $i$ (our instantaneous capacity $\kappa_{it}$), whereas the in-degree reflects the popularity of trader $i$, i.e. the number of people mirroring $i$, $P_{it}$. In Figure~\ref{supp:fig:data-label}, we present the distribution of $P_{it}$ across different instants. As we can see, the distribution of popularity of the different trades which follows a heavy tail distribution, a signature of preferential attachment given the strong impact of popularity on their mirroring decisions. More importantly, we see that the distribution is almost stable at different instants, a reflection of the dynamical stability of the system in the $\Omega$ period.

Using the {\tt parentCID} field from the trades dataset, we can identify whether a trade was executed independently or as a mirrored trade. If a trader has at least one trade with {\tt parentCID} $\neq 0$, we classify the trader as engaging in social learning. For statistical reasons, we restrict our analysis to traders who executed at least two trades during the observation window $\Omega$. The number of traders who meet this criterion is 164,634, of which 119,864 engage in mirror trading and 44,770 trade exclusively independently.

\subsection{Data Limitations}
Data was obtained from the eToro website. For that reason, all closed traders during the observation period were not recorded. The dataset was collected on a rolling basis, and there were periods when a missing connection led to entire days missing. There were about 30 such days of data missing, which accounts for 4\% 
is very small and should not substantially affect the results because the way we estimate popularity is relatively robust to missing data, and because the relative performance rankings of users should be similar after removing entire days.

In addition, approximately 1\% of traders are missing at random. We infer this from the presence of certain trades that are clearly copies, yet the corresponding original (parent) trades are not observed in the dataset. Notably, the available mimicked trades suggest that nearly all of these missing parent trades yielded either negligible or negative returns. Fortunately, since the replicated trades are observed, we are able to impute the missing parent trades based on this information.

The evidence of missing individual trades raises the possibility that additional trades from users who had no mimickers may also be missing without a trace. However, we can place an upper bound on the extent of missing data by examining trade IDs, which appear to be assigned sequentially. Excluding the 1,000 lowest trade IDs (likely corresponding to long-running trades initiated before our observation period), we observe approximately 92\% of the remaining possible trade IDs. Considering the 3\% of trades we already know are missing, this suggests that up to an additional 5\% of trades might be unobserved. Nevertheless, this is a loose upper bound; the true proportion may be much smaller if trade IDs are skipped for other reasons such as trades still open at the end of our observation window and therefore not appearing in the dataset of closed trades.

Another limitation of our dataset is the occasional presence of inconsistent values in certain columns. We identify 1,170 trades (approximately 0.001\%) with negative invested amounts, 182 trades (about 0.0002\%) where the recorded close date precedes the open date, and 6,470 trades (roughly 0.007\%) where the reported profit is inconsistent with the associated units invested and exchange rates. We attribute these anomalies to bugs or database errors within eToro’s system. Additionally, we are unable to fully reconstruct the precise relationship between the initially invested amounts and the number of units purchased. While our inferred relationship achieves less than 10\% relative error for 83\% of trades and less than 50\% error for 96\% of trades, the discrepancies suggest that key influencing variables—such as “stop loss” and “take profit” thresholds set by users—may be missing from the data we received. 

For more details about the data quality, see \cite{krafft2016rational} and \cite{Krafft2016HumanCollectiveIntelligence}.

\subsection{Computing the performance and popularity of traders} \label{supp:data:pop_per}
From the mirror dataset, we can compute the popularity of each trader at a given time $t$. A trader $i$ has a popularity $P_{it}$ at time $t$, defined as the number of mirroring relationships where $i$ is the {\tt parentCID} and  {\tt openDate} $< t <$ {\tt closeDate} holds.

From the trades data set we can compute the performance for each individual trade
$$\rho_{it} = \frac{\tt{nProfit}_{it}}{\tt{unitsDecimal}_{it} \times \tt{openRate}_{it}}.$$
Note that here $t$ refers to the {\tt closeDate} of a trade, which is recorded as a timestamp with second-level precision. The 30-day rolling performance of trader $i$, denoted $R_{it}$, is computed as:
\begin{equation} 
\label{supp:data:rollingperformance}
    R_{it} = \frac{1}{N_{it}} \sum_{\tau \in [t- \Delta, t]} \rho_{i\tau},
\end{equation}
and  $\rho_{i\tau}$ is the performance of a trade closed by trader $i$ at time $\tau$, $N_{it}$ is the number of trades trader $i$ closed in the interval $[t - \Delta, t]$, and $\Delta$ corresponds to a 30-day duration. 

\begin{figure}[t] 
    \centering
    \includegraphics[width = 10 cm]{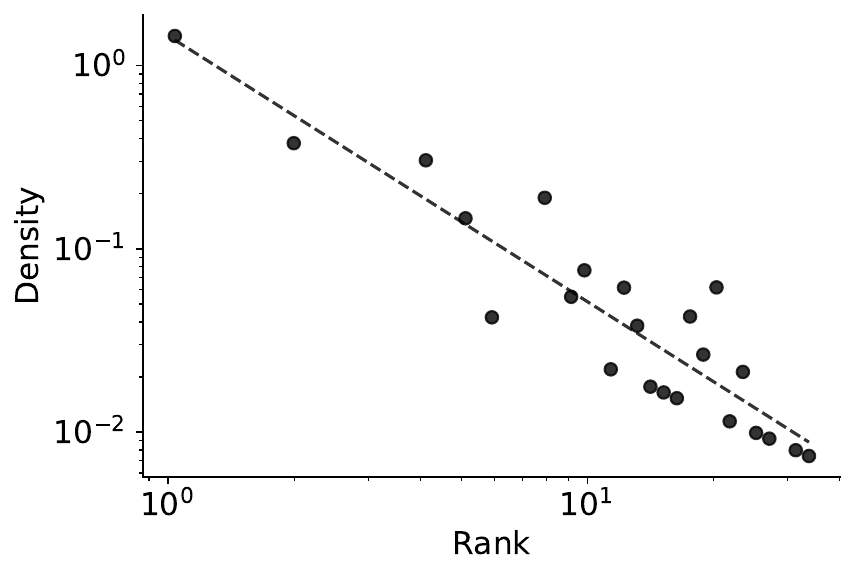}
    \caption{\textbf{Density of edges removed as a function of their rank.} Users terminate a mirroring connection depending on their rank according to Equation~\ref{supp:eq:rank} . The lower-ranked mirrors have a higher probability of getting terminated. Line shows a Zipf-law fit to the density as $P(\text{rank}) \sim \text{rank}^{-\beta}$ with $\beta = 1.45 \pm 0.13$.}
    \label{supp:fig:rank}
\end{figure}

\section{Analysis of Mirroring Removal} 
\label{sup:sec:noderemoval}
From the analysis of $\eta_i^+$, $\eta_i^-$, and $\kappa_i$, we conclude that there is a correlation between $\eta_i^+$ and $\eta_i^-$, and that traders tend to maintain a relatively constant capacity $\kappa_i$. As a result, when a new mirror is added, the trader must remove an existing one. To understand the dynamics of how traders remove their existing mirrors, we study whether they use similar information as when they created them. Specifically, at time $t$ and for trader $i$ we rank their mirrors inversely proportionally to a scoring function similar to the one used to create mirrors:
\begin{equation} \label{supp:eq:rank}
P(i \rightarrow j; t) \simeq \mathrm{logit}^{-1}\left[ \beta_{\text{perf}} R_{jt} + \beta_{\text{pop}} P_{jt} \right],
\end{equation}
Thus, mirrors with lower ranking are the ones that have the lowest $P(i \rightarrow j;t)$. For statistical reasons, we restrict our analysis to traders who had more than 10 active mirrors at time $t$. As we can see Figure~\ref{supp:fig:rank}, most of the mirrors removed are those with lower ranking ($\sim 50$\% with rank 3 or lower), suggesting that traders use similar utility functions to evaluate creating and removing mirrors.

\section{Modeling dynamical social learning}\label{supp:sec:model}
The interplay between informational factors—popularity and performance—and the limited-capacity, dynamic strategies adopted by traders is central to understanding behavior on social learning platforms. To investigate that interplay dynamics, we propose a model of social learning informed by patterns observed in the eToro dataset. The platform is modeled as a directed temporal network, where nodes represent traders and edges capture mirroring relationships, directed from follower $i$ to mirrored trader $j$. 

Each trader, $i$, is characterized by a dynamic mirroring strategy $(\kappa_i, \eta_i^+)$, where $\eta_i^+$ governs the probability of initiating a new mirroring link at each time step. In our model, $\kappa_i$ is drawn from a Poisson distribution with mean $10$ and $\eta_i^+$ is defined as a uniformly distributed ratio in $[0,1]$, sampled independently for each agent. Those distributions mimic the distributions of and absence of correlation between $\eta_i^+$ and $\kappa_i$ in the eToro platform (see Figure 2 in the main paper).

At each time step, the daily return for agent $i$ at time $t$ is given by:
\begin{equation}
\rho_{it} = \frac{1}{\kappa_{i,t-1}} \sum_{j} A_{ij,t-1} \rho_{j,t-1}  + \epsilon_{it},
\end{equation}
where $A_{ij, t-1}$ is the element of the adjacency matrix indicating a mirroring relationship from trader $i$ to trader $j$, and $\kappa_{i,t-1} = \sum_j A_{ij,t-1}$ denotes the number of traders mirrored by trader $i$ at time $t-1$ (our instantaneous individual capacity at time $t-1$). We model the independent investment component, $\epsilon_{it}$, using an Ornstein–Uhlenbeck (OU) process. We scale $\rho_{it}$ by a factor of 0.5 to ensure equal weighting between mirrored investments and the agent’s own independent investments.  Furthermore, each trader who mirrors trader $j$ receives the full return $\rho_{j,t-1}$, meaning all followers obtain the same mirrored signal from $j$ regardless of how many followers $j$ has.

\begin{figure}[t]
\centering

\begin{minipage}{.5\textwidth}
  \centering
  \makebox[-210pt][r]{\raisebox{4.4em}{\hspace{6.6cm}\textsf{\textbf{\LARGE A}}}}  


\captionof*{table}{Power-law fits for different $(\beta_{\text{pop}}, \beta_{\text{per}})$ pairs.}
\resizebox{\linewidth}{!}{
\begin{tabular}{lcccc}
\toprule
$(\beta_{\text{pop}}, \beta_{\text{per}})$ & (0.34, 15.68) & (15.68, 15.68) & (10.00, 20.00) & (20.00, 10.00) \\
\midrule
$\beta$         & 1.51*** & 1.52***  & 1.48***  & 1.51***  \\
                  & (0.01) & (0.01)  & (0.01) & (0.01) \\
$x_{\min}$        & 2.00     & 2.00     & 2.00     & 2.00     \\
\hline 
$\mathrm{Observations}$  & 1515309   & 774151    & 896437    & 711769    \\
$\log L_{\mathrm{PL}}$& -5335872.591 & -3055994.643 & -3429196.865 & -2641671.631 \\
$\log L_{\mathrm{LN}}$   & -5324724.706 & -3051240.225 & -3424624.701 & -2640743.267 \\
$R$                      & -11147.885   & -4754.418    & -4572.164    & -928.364     \\
\bottomrule
\end{tabular}
}
\vspace{0.5em}
\noindent\footnotesize\textit{Note:} $^{*}$p$<$0.1; $^{**}$p$<$0.05; $^{***}$p$<$0.01.
\label{supp:table:degree-model}
\vspace{3.5em}
\end{minipage}%
%
\begin{minipage}{.5\textwidth}
  \centering
  \makebox[-210pt][r]{\raisebox{-1.5em}{\hspace{6.6cm}\textsf{\textbf{\LARGE B}}}}  

  \vspace{1.5em} 
  \includegraphics[width=\linewidth]{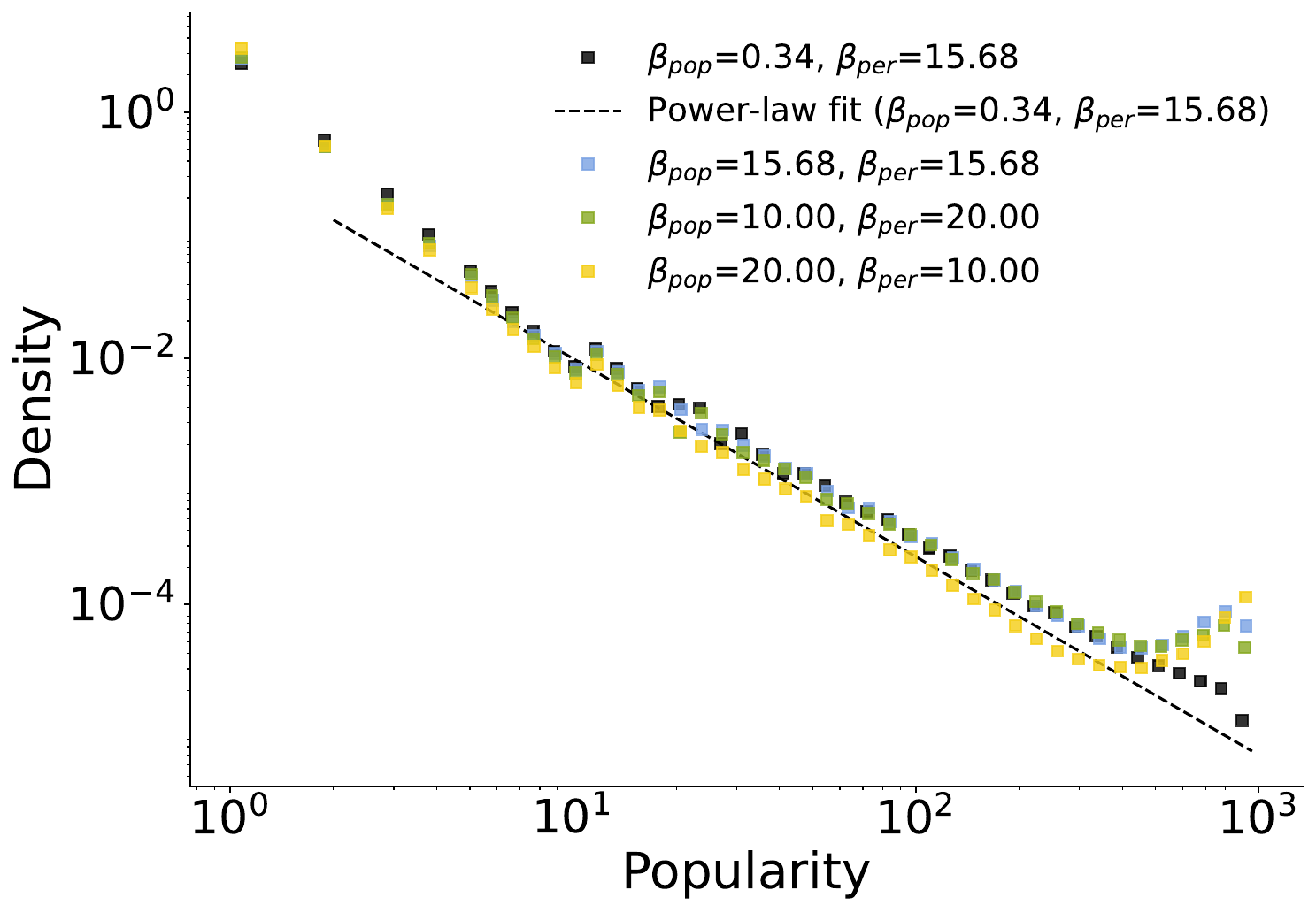}
\end{minipage}

\caption{\textbf{Distribution of popularity in the model.} A: Popularity distribution in the model fitted to a power-law model, $ P(P_{it}) = A P_{it}^{-\beta}$ using the package {\tt powerlaw} \cite{power-law}. The table reports the estimated coefficients and their corresponding standard errors (in parentheses). The p-values are based on two-sided tests used to evaluate the hypothesis that each coefficient is significantly different from zero. The $\log L_{pl}$ is the log-likelihood of the data under a power-law distribution, $\log L_{ln}$ is the log-likelihood of the data under a log-normal distribution and $R = \log L_{pl} - \log L_{ln}$ is the likelihood ratio. B: The graphical representation of popularity distribution in the model for different $\beta_{\text{per}}$ and $\beta_{\text{pop}}$.}
\label{supp:fig:general-label}
\end{figure}

It is important to note that in our model, a trader $i$’s performance at time $t$ depends on the performance of her mirrored traders at time $t-1$. We use $t-1$ rather than $t$ to simplify the model and to prevent feedback loops in the network that could artificially inflate $\rho_{it}$. For example, if trader A mirrors B and B mirrors A, or in more complex cycles such as A $\rightarrow$ B $\rightarrow$ C $\rightarrow$ A, using the same time step would introduce recursive dependencies and potentially lead to spurious amplification artifacts.

During each iteration, a trader decides whether to revise her mirroring relationships based on her individual mirroring propensity $\eta_i^+$. If a trader decides to modify her mirrors, she selects a new trader $j$ to follow with a probability given by:
\begin{equation} 
P(i \rightarrow j; t) \simeq \mathrm{logit}^{-1}\left[ \beta_{\text{perf}} R_{jt} + \beta_{\text{pop}} P_{jt} \right],
\end{equation}
where $R_{jt}$ is the 30-day rolling performance and $P_{jt}$ is the popularity of trader $j$. When a trader adds a new mirrored connection, she must also drop one. The mirror with the lowest $P(i \rightarrow j)$ is removed. The model parameters $\beta_{\text{per}}$ and $\beta_{\text{pop}}$ are used to adjust the emphasis node $i$ applies on $R_{jt}$ and $P_{jt}$. It is worth noting that in this equation, we have omitted the constants present in Equation~\ref{eqLogistic} from the main text, as they are not necessary for our model.
\begin{figure}[t] 
    \centering
    \includegraphics[width = \linewidth]{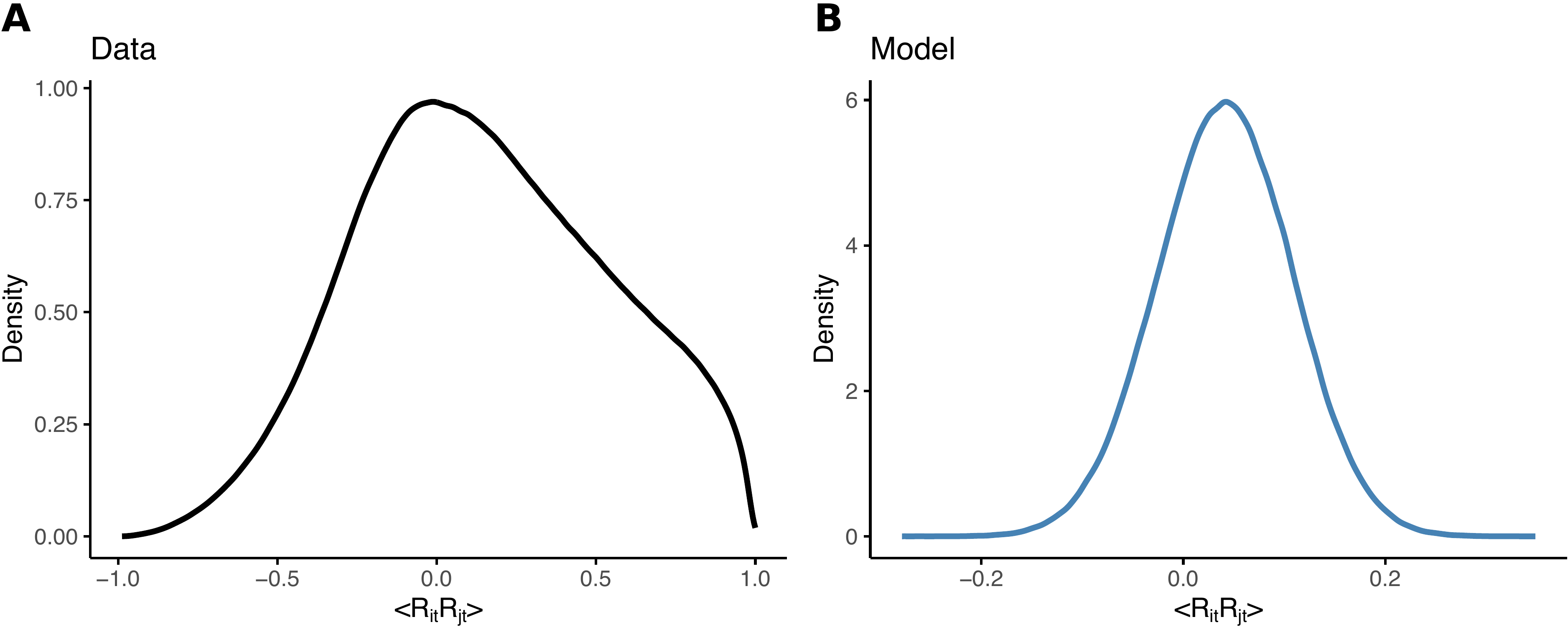}
    \caption{\textbf{Correlation between performance of different traders.} A: Correlation of $R_{it}$ between different traders in the eToro data. B: Correlation of $R_{it}$ between different traders in the model where $\beta_{\text{per}} = 0.34$ and $\beta_{\text{pop}} = 15.68$. }
    \label{supp:fig:Correlation}
\end{figure}

\subsection{Ornstein--Uhlenbeck Process}\label{supp:data:OU}
The Ornstein–Uhlenbeck (OU) process governing user $i$'s independent returns at time \( t \) is given by:
\begin{equation}
d\epsilon_{it} = \theta (\mu - \epsilon_{it})\,dt + \sigma\,dW_{it},
\end{equation}
where \( \epsilon_{it} \) denotes the independent returns, \( \mu \) is the long-term mean, \( \theta \) is the rate of mean reversion, \( \sigma = \sqrt{2\theta} \) is the volatility (standard deviation of the process), and \( W_{it} \) is a standard Wiener process. To approximate the empirical patterns observed on eToro—where the average return of users is \( -10.98 \pm 0.41 \, \text{bps} \), and the correlation between individual average performance and \( \gamma_i \) is \( 2.06 \pm 0.28 \), we set \( \mu = -0.1 \, \text{bps} \) and \( \theta = 0.65 \). With these parameters, and by fixing \( \beta_{\text{per}} = 0.34 \) and \( \beta_{\text{pop}} = 15.68 \), the simulation produces an average return of \( \overline{\langle \rho_{it} \rangle}_t = -16.97 \pm 1.81 \, \text{bps} \), and a correlation between individual average performance and \( \gamma_i \) of \( 1.6 \pm 0.43 \) (see regression results in Figure~\ref{supp:table:S5_regression_panelC}), which reasonably approximates the empirical value. These parameters also produce a cross-correlation between \( R_{it} \) and \( P_{it} \) that closely resembles what is observed in the empirical data (see Figure~\ref{supp:fig:ccf}). Note that those parameters are not fit to exactly match the values in the empirical data, as our model is intentionally minimal and assumes all agents share the same individual trading dynamics. This simplifying assumption allows us to isolate and understand the effects of social learning mechanisms without introducing heterogeneity in agent behavior or optimizing parameter fits.

\subsection{Degree distribution of the model}
In the simulated model, we assign each trader a capacity $\kappa$, which is its out-degree and therefore the distribution of the out-degree is pre-determined. However, the distribution of their popularity (in-degree) is not pre-determined and will depend on the model dynamics. To compare our model with eToro, we set $\beta_{\text{per}}$ and $\beta_{\text{pop}}$ to $0.34$ and $15.68$. From Figure~\ref{supp:fig:general-label}, we can see the exponent doesn't change much with time and closely matches that of the eToro platform shown in Figure~\ref{supp:fig:data-label}. This result suggests that our model can capture the popularity dynamics in the social trading platform in an accurate way.

\section{Dynamic Properties of Popularity and Performance }
\subsection{Correlation between traders' performance}\label{sup:subsec:corrR}
In this section, we compare the correlation of $R_{it}$ between different traders. We do this both for the real data and our model.
In Figure~\ref{supp:fig:Correlation} A, we compute $R_{it}$ of traders as a time series and then we compute the correlation between the $R_{it}$ of different traders that have traded for at least $50$ days during $\Omega$. The mean of the distribution is $0.07 \pm 0.01$. In Figure~\ref{supp:fig:Correlation} B, we have plotted the density function of the correlation of $R_{it}$ between different traders in the model. In the model used to generate the plot, we have used $\beta_{\text{{per}}} = 0.34$ and $\beta_{\text{pop}} = 15.68$. The mean of the distribution is at $0.04 \pm 0.01$. Even though we used the OU process to simulate individual investments in our model, the mirroring relationship leads to a positive correlation between the users. Since results from subfigures A and B in Figure~\ref{supp:fig:Correlation} are very similar to each other, this suggests that our model, despite its simplicity, is effective.

\subsection{Correlation of traders popularity and performances} \label{supp:subsec:correlatonPR}
\begin{figure}[t] 
    \centering
    \includegraphics[width = \linewidth]{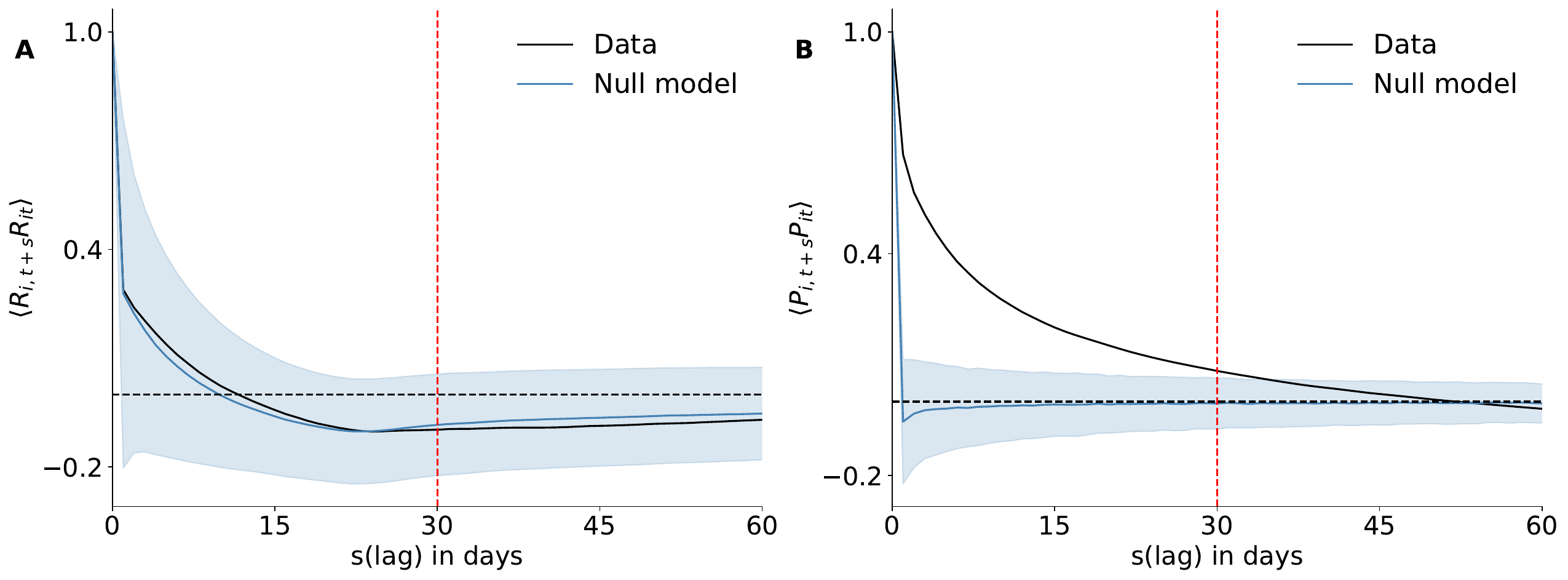}
    \caption{\textbf{Autocorrelation of performance and popularity.} A: Autocorrelation of 30-day rolling performance. The blue shaded region indicates the standard deviation of the mean calculated from the null model. B: Autocorrelation in popularity. The blue shaded region indicates the standard deviation of the mean calculated from the null model.  }
    \label{supp:fig:Auto-Correlation}
\end{figure}

From our dataset, we compute both the popularity, $P_{it}$, and the 30-day rolling performance, $R_{it}$, of various traders over different time periods, and analyze their autocorrelation up to a time lag of 60 days. These results are presented in Figure~\ref{supp:fig:Auto-Correlation}. We compare the real data with a null model in which we shuffle the daily performance or the popularity of each user across the time period $\Omega$. We recompute $R_{it}$ using Equation~\ref{supp:data:rollingperformance} for the shuffled data.

As we see, the autocorrelation of $R_{it}$ closely matches the null model, indicating that any observed autocorrelation is primarily an artifact of the rolling window. This aligns with the intuition that a trader’s performance on a given day does not predict their future performance. In contrast, the autocorrelation in popularity deviates substantially from the null model, suggesting that popularity evolves more slowly over time compared to $R_{it}$. 

\begin{figure}[t] 
    \centering
    \includegraphics[width = 0.8\linewidth]{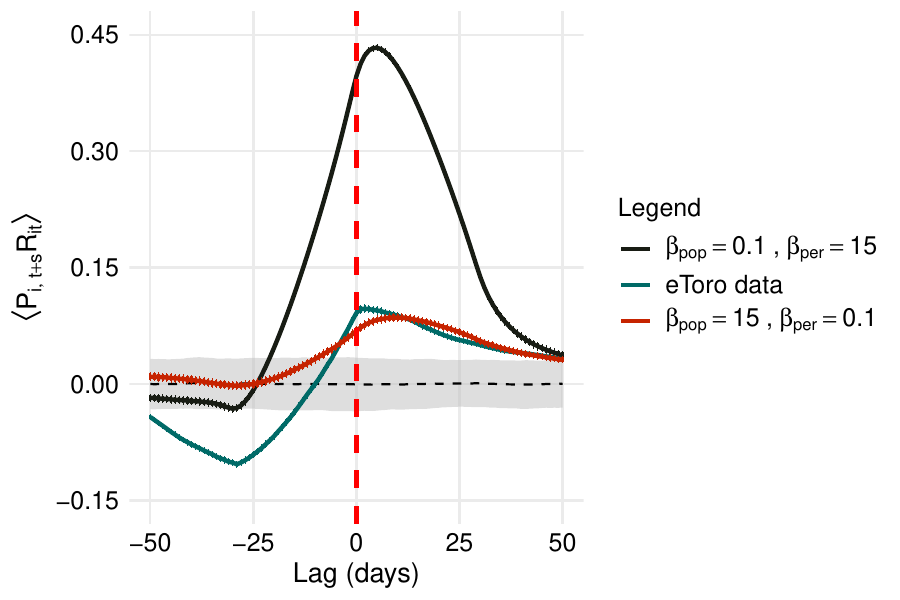}
    \caption{\textbf{Cross correlation between popularity and performance.} The cross-correlation between 30-day rolling performance, $R_i$, and popularity, $P_i$ — denoted as $\left\langle P_{i,t+s} , R_{it} \right\rangle$ — is averaged across all traders $i$. The green line represents the real data from the eToro platform, whereas the black and red lines correspond to different $\beta_{\text{perf}}$ and $\beta_{\text{pop}}$ values in the model.}
    \label{supp:fig:ccf}
\end{figure}

\subsection{Cross-correlation between popularity and performance} \label{supp:subsec:cross-correlation}
 
We compute the cross-correlation between a trader’s performance $R_{it}$ and their future popularity $P_{i,t+s}$, denoted as $\langle P_{i,t+s} \, R_{it} \rangle$, for each trader on the eToro platform as well as in our model. In Figure~\ref{supp:fig:ccf}, the bold green line represents the empirical cross-correlation observed on eToro. We observe that a trader's performance at time $t$ is positively correlated with their popularity at future times $t+s$, although the correlation is small. 

We perform the same analysis for traders in our model. In Figure~\ref{supp:fig:ccf}, the black line corresponds to the case where traders place greater emphasis on performance than popularity when making mirroring decisions ($\beta_{\text{perf}} = 15$, $\beta_{\text{pop}} = 0.1$), while the red line represents the opposite scenario, where popularity is prioritized over performance ($\beta_{\text{per}} = 0.1$, $\beta_{\text{pop}} = 15$).

These results illustrate that the cross-correlation between $R_{it}$ and $P_{i,t+s}$ strengthens when traders focus more on performance than on popularity. In such cases, performance becomes a predictor of future popularity, highlighting the causal influence of performance-driven mirroring decisions on popularity growth.

\section{Libraries used}
Analysis was conducted in Python and R using the following packages:
\begin{itemize}
    \item Logistic and least square regressions with fixed effects were done using the {\tt fixest} library \cite{fixest}.
    \item Network analysis was carried out using the {\tt networkX} package \cite{hagberg2008exploring}.
    \item The power law fit was done using the Python package {\tt powerlaw} \cite{power-law}
    \item Ordinary least squares regression was also performed using {\tt scipy.stats.linregress} from the {\tt SciPy} package \cite{Virtanen2020SciPy}.
    \item Package {\tt matplotlib}  was used for the visualizations \cite{Hunter:2007}.
\end{itemize}

\begin{table}[ht]
\centering
\resizebox{\textwidth}{!}{%
\begin{tabular}{lccccccc}
\toprule
 $(\beta_{\text{pop}}{,} \beta_{\text{per}})$ 
   & $(20.00, 0.00)$ & $(15.00, 5.00)$ & $(10.00, 10.00)$ 
   & $(15.68, 0.34)$ & $(5.00, 15.00)$ & $(0.00, 20.00)$ 
   & etoro $(15.68, 0.34)$\\
\midrule
Slope 
  & -0.01      & 27.45*** & 76.05*** & 1.65**   & 129.75*** & 171.32*** & 2.05*** \\
     & (0.25)     & (2.93)   & (10.35)  & (0.42)   & (20.89)   & (29.79)   & (0.28)  \\
Intercept
  & -10.83***  & 97.47*** & 243.95***& 0.49     & 411.02*** & 559.39*** & 11.45***\\
     & (0.23)     & (2.69)   & (9.51)   & (0.39)   & (19.17)   & (27.32)   & (1.63)  \\
\midrule
Observations & 997,894  & 997,884 & 997,869 & 997,906 & 997,885 & 997,902 & 3,359 \\
Squared Corr & 0.00     & 0.92    & 0.87    & 0.65    & 0.83    & 0.81    & 0.86  \\
$R^2$        & 0.00     & 0.92    & 0.87    & 0.65    & 0.83    & 0.81    & 0.86  \\
\bottomrule
\end{tabular}%
}
\caption{Linear‐log regression results for Figure~\ref{fig:heatmap}C in the main text. The $p$‐values are based on two‐sided tests used to evaluate whether each coefficient differs from zero. Standard errors are in parentheses. All curve fits performed with \texttt{scipy.stats.linregress} (SciPy).}
\label{supp:table:S5_regression_panelC}
\end{table}

\small

\end{document}